\documentclass{amsproc}

\usepackage[cmtip,all]{xy}

\usepackage{amsmath,epsf,amssymb,latexsym,amsthm,cite,setspace,bbm, amsfonts}

\DeclareFontFamily{U}{rsf}{}
\DeclareFontShape{U}{rsf}{m}{n}{
  <5> <6> rsfs5 <7> <8> <9> rsfs7 <10-> rsfs10}{}
\DeclareMathAlphabet\Scr{U}{rsf}{m}{n}

\def\AC#1#2{{\{#1,#2\}}}

\def\cDb{{\overline{\cD}}}
\def\cQb{{\overline{\cQ}}}

\def\C{{\mathbb C}}

\def\P{{\mathbb P}}
\def\R{{\mathbb R}}
\def\Z{{\mathbb Z}}

\def\End{\operatorname{End}}

\def\Pic{\operatorname{Pic}}

\def\SO{\operatorname{SO}}

\def\GU{\operatorname{U{}}}

\def\GE{\operatorname{E}}

\def\p{\partial}
\def\pb{\bar{\partial}}

\def\ra{\rangle}

\def\ff#1#2{{\textstyle\frac{#1}{#2}}}
\def\half{\frac{1}{2}}

\def\cD{{\cal D}}

\def\cG{{\cal G}}
\def\cH{{\cal H}}

\def\cJ{{\cal J}}

\def\cL{{\cal L}}

\def\cO{{\cal O}}

\def\cQ{{\cal Q}}

\def\cT{{\cal T}}

\def\cX{{\cal X}}
\def\cY{{\cal Y}}

\newcommand\alphab{\overline{\alpha}}
\newcommand\betab{\overline{\beta}}
\newcommand\gammab{\overline{\gamma}}

\newcommand\etab{\overline{\eta}}
\newcommand\thetab{\overline{\theta}}

\newcommand\chib{\overline{\chi}}

\newcommand\nut{\widetilde{\nu}}

\newcommand\cb{\overline{c}}

\newcommand\yb{\overline{y}}
\newcommand\zb{\overline{z}}

\newcommand\Lb{\overline{L}}

\newcommand\Wb{\overline{W}}

\newcommand\Yb{\overline{Y}}

\copyrightinfo{2014}{American Mathematical Society}

\def\br{{\boldsymbol{r}}}

\def\bK{{{\boldsymbol{K}}}}

\def\bY{{{\boldsymbol{Y}}}}

\def\lcm{\operatorname{lcm}}
\def\coker{\operatorname{coker}}

\def\cXb{{{\overline{\cX}}}}
\def\cYb{{{\overline{\cY}}}}

\def\bQ{{{{\boldsymbol{Q}}}}}
\def\bQb{{{\overline{\bQ}}}}

\def\Sym{\operatorname{Sym}}
\def\bq{{{{\boldsymbol{q}}}}}
\def\bqb{{{\overline{\bq}}}}

\def\pz{\partial_z}
\def\pbz{\overline{\partial}_{\zb}}
\def\Dz{D_z}

\def\cal#1{\mathcal#1}

\def\htt{\tilde h}

\newtheorem{theorem}{Theorem}[section]

\theoremstyle{definition}

\theoremstyle{remark}

\numberwithin{equation}{section}

\begin{document}

\title{Massless spectrum for hybrid CFTs}


\author{Marco Bertolini}
\address{Center for Geometry and Theoretical Physics, Box 90318, Duke University, Durham, NC 27708-0318}
\email{mb266@phy.duke.edu}

\author{Ilarion V.~ Melnikov}
\address{George P. and Cynthia W. Mitchell Institute for Fundamental Physics and
Astronomy, Texas A\&M University
College Station, TX 77843, USA}
\email{ilarion@physics.tamu.edu}

\author{M.~Ronen Plesser}
\address{Center for Geometry and Theoretical Physics, Box 90318, Duke University, Durham, NC 27708-0318}
\email{plesser@cgtp.duke.edu}


\date{\today}

\begin{abstract}

We describe a class of theories obtained by fibering a Landau-Ginburg orbifold over a compact K\"ahler base. While such theories are often 
described as phases of some GLSM, our description is independent of such an embedding. We provide a method for computing the massless spectrum.
This note is based on the longer paper arXiv:1307.7063.

\end{abstract}

\maketitle


\section{Introduction}

Elucidating the structure of the moduli space of $(0,2)$ heterotic compactifications is a intriguing problem but far from solved.
A useful tool which has been fruitfully employed to explore this area is the gauged linear sigma model (GLSM) \cite{Witten:1993yc}. 
In this context, non linear sigma models (NLSM)
on compact Calabi-Yau manifolds and Landau-Ginzburg (LG) theories describe low-energy dynamics for special limiting values of the parameters.
A GLSM expert has surely already encountered a model exhibiting a hybrid phase, which in simple words consists of a LG orbifold (LGO) fibered non-trivially
over a compact base. We provide an intrinsic definition of a hybrid SCFT, i.e. independent of a GLSM embedding and its UV completion,
in principle describing a new class of heterotic vacua.

As is well known, gauge neutral massless states in spacetime correspond to first order deformations of the internal SCFT. 
While there exist mechanisms \cite{Beasley:2003fx,Silverstein:1995re} that prevent a subset of these deformations from being lifted by worldsheet instantons, there are
situations \cite{Aspinwall:2011us} where instantons give masses to singlets. Moreover, the authors in \cite{Aspinwall:2010ve} started a more systematic study of 
the behavior across phases of massless singlets corresponding in the large radius limit to $h^1(\End T)$\footnote{These are the bundle moduli in a (2,2) exclusive language.}
in examples with $(2,2)$ worldsheet supersymmetry realized as hypersurfaces in toric varieties. 
By extending the techniques for NLSM and LGO \cite{Kachru:1993pg}, we provide a method to compute the 
massless spectrum of hybrid models in the hybrid limit, where the base manifold is taken to be large. 
Such a technique is particularly useful for at least two reasons: on the one hand
we increase the number of points/limits where exact computations can be carried by a fair amount; on the other hand, one can hope to tackle the computation 
of worldsheet instantons more easily in a hybrid set-up, since one has to deal with rational curves on the lower dimensional base instead of a Calabi-Yau three-fold.

\section{Hybrid geometric set-up}

Our goal is to construct a non-trivial SCFT obtained by hybrid compactification. 
The starting point for our model is a NLSM\footnote{We will work with a flat worldsheet.} on a K\"ahler manifold $Y_0$ together with a holomorphic function $W: Y_0 \rightarrow \C$ such that 
$dW^{-1}(0)=B\subset Y_0$, where $B$ is compact and K\"ahler. This {\it potential condition} will turn out to be fairly important in what follows.
This geometry will realize a {\it hybrid model} when locally for $B\subset Y_0$ it can be modeled as the  total space of a rank $n$ holomorphic vector bundle, $\bY : X \rightarrow B$.
As we will see, this is quite natural since the bosonic potential of the hybrid action will suppress finite fluctuations supported away from $B$.

\subsection*{Action, symmetries and the hybrid limit}

In this section we are going to construct an action for a hybrid model and analyze its symmetries. We will work in $(2,2)$ superspace with coordinates $(z,\zb,\theta,\thetab,\theta',\thetab')$ and we define the supercharges 
\begin{align}
\cQ = -{\p\over\p\theta} + \thetab \p~, \quad \cQb= -{\p\over\p\thetab} + \theta \p~, \quad
\cQ' = -{\p\over\p\theta'} + \thetab' \pb~, \quad \cQb'= -{\p\over\p\thetab'} + \theta' \pb~, 
\end{align}
and the superderivatives 
\begin{align}
\cD = {\p\over\p\theta} + \thetab \p~, \quad \cDb= {\p\over\p\thetab} + \theta \p~,\quad
\cD' = {\p\over\p\theta'} + \thetab' \pb~, \quad \cDb'= {\p\over\p\thetab'} + \theta' \pb~, 
\end{align}
where $\p \equiv \p / \p z$ and similarly for $\pb$. The non-trivial anti-commutation relations are
\begin{align}
\{\cD,\cDb\}=2\pb~,\qquad \{\cQ,\cQb\}=-2\pb~,
\end{align}
and similarly for the primed quantities.  
These objects are graded by the $R$-symmetry $\GU(1)_L(\GU(1)_R)$, which assign value $+1$ to $\theta'(\theta)$ and $-1$ to $\thetab'(\thetab)$ respectively.

The building blocks for a non-compact NLSM on $\bY$ are the (2,2) bosonic chiral (and their conjugate anti-chiral) superfields 
\begin{align}
\label{eq:02sfields}
\cY^\alpha &= Y^\alpha + \sqrt{2} \theta' \cX^\alpha + \theta'\thetab' \pz Y^\alpha~, &&&
\cYb^{\alphab} & = \Yb^{\alphab} - \sqrt{2}\thetab' \cXb^{\alphab} -\theta'\thetab' \pz \Yb^{\alphab}~.
\end{align}
The lowest components of the (0,2) bosonic chiral multiplets $Y^\alpha$ are coordinates on $\bY$, which can be split into fiber coordinates, indicated as $\phi^i$, $i=1,\dots,n$, and
base coordinates, $y^I$, $I=1,\dots,d$. The lowest components of the chiral fermi multiplets $\cX^\alpha$ are left-moving fermions $\chi^\alpha$. While in general the $\chi^\alpha$
couple to a stable holomorphic bundle $E\rightarrow \bY$ subject to the conditions
\begin{align}
c_1(E)=0~,\qquad c_2(E)=c_2(T_{\bY})~,
\end{align} 
for the purpose of this paper we will assume they couple to the tangent bundle. 

The $(2,2)$ action for our hybrid model is given by
\begin{align}
\label{eq:22action}
S &= \frac{1}{8\pi} \int d^2 z ~\cD \cDb \cDb' \cD' \bK(\cY,\cYb) + \frac{m}{4\pi} \int d^2z ~\cD\cD' W(\cY) + \text{c.c.}~.
\end{align}
It is easy to recognize the two terms: the first is a term kinetic term realizing a NLSM on $\bY$ while the second is a superpotentiatial term, which satisfies the potential condition $d W(0)^{-1}=B$. 
This implies that the bosonic potential is minimized by $B$ and the low energy physics is described by small fluctuations around $B$.
Integrating over the fermionic coordinates and eliminating the auxiliary fields by means of their equations of motion we obtain the component action, which after some work it reads
\begin{align}
\label{eq:compaction}
2\pi L & = \rho_\alpha \pbz y^\alpha + \chib_\alpha \pbz \chi^\alpha + \eta^\alpha\left[ g_{\alpha\betab} \Dz \etab^{\betab} + \etab^{\betab} R_{\alpha\betab\gamma}^{~~~~\delta} \chib_\delta\chi^\gamma +\chi^\beta D_\alpha W_\beta
\right]~
\nonumber\\
&\qquad+\chib^{\alphab}\etab^{\betab} D_{\betab}\Wb_{\alphab} + g^{\betab\alpha} W_\alpha \Wb_{\betab}~,
\end{align}
where we have implemented the following field redefinitions
\begin{align}
\chib_\alpha = g_{\alpha\beta}\chib^{\betab}~,\qquad \rho_\alpha = g_{\alpha\alphab}\p \yb^{\alphab} + \Gamma^\delta_{\alpha\gamma}\chib_\delta \chi^\gamma~.
\end{align}
The implications of the non-covariant transformation property of $\rho$ will become clear below.

Similarly to the LGO case, the $\GU(1)_L\times\GU(1)_R$ symmetries of the $W=0$ theory introduced above play a key role in constructing a heterotic/Type II vacuum. 
In fact, they are necessary to define left/right spectral flow operators, whose consequences are modular invariance and space-time supersymmetry. 
However, this symmetry is generically broken by the superpotential. We then demand the existence of a holomorphic Killing vector $V$ 
such that $\cL_V W= W$. The action of $V$ on the fields, which we indicate as $\delta$, is non-chiral and it is easy to check that 
$\delta_{L,R}\equiv \delta_{L,R}^{\text{old}}+\delta$ are actual symmetries of the classical action.

What about the quantum theory? 
The action of $\GU(1)_L$ is chiral and it can suffer from anomalies\footnote{The action of $\GU(1)_{\text{diag}}\subset\GU(1)_L\times\GU(1)_R$ is non-chiral, and it will not be anomalous.}. 
The anomaly vanishes if $c_1(T_{\bY})=0$, that is 
 if $\bY$ is a Calabi-Yau manifold, which we will assume for the rest of the paper. 
More specifically, we will assume that the canonical bundle $\cO_{\bY}$ is trivial.

So far we have discussed symmetries of the classical and quantum UV theory. In order to be able to identify these symmetries with those in the corresponding IR SCFT we expect 
our hybrid model to flow to, 
we need to impose one additional constraint on $V$. We require $V$ to be a vertical vector field, that is $\GU(1)_L\times\GU(1)_R$ fix $B$ point-wise. 
More precisely, we have that $\cL_V \pi^\ast(\omega)=0$ for each $\omega\in\Omega^\bullet(B)$. A model 
satisfying this additional constraint is denoted a {\it good hybrid}. 

The action is by construction $(2,2)$ supersymmetric, but for our subsequent goal of describing the massless spectrum 
we focus on the action of $\bQb=\bQb_0 + \bQb_W$, where $\bQb_0$ is the supercharge of
the non-compact NLSM of the $W=0$ theory while $\bQb_W$ contains all the superpotential dependence. 
Moreover, they satisfy $\bQb_0^2=\bQb_W^2=\AC{\bQb_0}{\bQb_W}=0$.
Up to the equations of motion, we find that the non-trivial actions are 
\begin{align}
\label{eq:supersplit}
\bQb_0 \cdot \yb^{\alphab} &= -\etab^{\alphab}~,&
\bQb_0 \cdot \eta^\alpha &= \pbz y^\alpha~,&
\bQb_W \cdot \chib_\alpha & = W_\alpha~,&
\bQb_W \cdot \rho_\alpha & = \chi^\beta  W_{\beta\alpha}~. 
\end{align}

In general one might expect the vacuum to be destabilized by worldsheet instantons, that is non-trivial maps wrapping rational curves on the base. 
In analogy to the ``large radius limit" for a compact NLSM, we define the {\it hybrid limit} as the K\"ahler class of $B$ is taken deep into the K\"ahler cone.
In other words, $B$ is large and we expand around trivial maps.

Now we turn to a technique for computing the massless spectrum of a hybrid model in the hybrid limit.

\section{Massless spectrum of a hybrid model}

\subsection*{Space-time generalities}

We briefly review the procedure to obtain a critical heterotic theory starting with our hybrid $c=\cb=9$ $N=(2,2)$ internal SCFT.
Following Gepner's prescription, we add $10$ left moving fermions $\lambda_A$ which realize a $SO(10)$ level one current algebra and a (hidden) level one $\GE_8$ current algebra, in addition to the degrees of freedom of the uncompactified spacetime $\R^{1,3}$.
Modular invariance is enforced by separate GSO projections on both left- and right-moving fermion numbers.
In addition, the  left-moving GSO projection enhances the space-time gauge group from the linearly realized $\SO(10)\times\GU(1)_L$ to $\GE_6$,
while the right-moving GSO projection ensures $N=1$ space-time supersymmetry.
In particular, space-time supersymmetry implies that the knowledge of the massless fermion spectrum, which is identified with the right-moving Ramond ground states, is sufficient for determining the full massless space-time spectrum.

On-shell massless fermions will satisfy the conditions $\Lb_0=L_0=0$, where $\Lb_0$ and $L_0$ are the right- and left-moving energies in the internal theory.
The $\GU(1)_L$ charge $\bq$ will determine the corresponding $\GE_6$ representation according to the decomposition
\begin{align}\label{E6decomposition}
&\GE_6 \mapsto \SO(10)\times \GU(1)_L \nonumber\\
& \mathbf{78} \mapsto \mathbf{45}_0 + \mathbf{16}_{-3/2} +  \mathbf{\overline{16}}_{3/2}  + \mathbf{1}_0    \nonumber\\
& \mathbf{27} \mapsto  \mathbf{16}_{1/2} +  \mathbf{{10}}_{-1}  + \mathbf{1}_2    \nonumber\\
& \mathbf{\overline{27}} \mapsto  \mathbf{\overline{16}}_{-1/2} +  \mathbf{{10}}_1  + \mathbf{1}_{-2} .
\end{align}
The type of space-time multiplet will be determined by the following simple rule: states with $\bqb=-\half$ and $\bqb=\half$ belong to chiral and antichiral multiplets respectively, 
while states with $\bqb=\pm\ff{3}{2}$ belong to vector multiplets.

As in the more familiar set-up of LGOs, it is convenient to combine the GSO projection with the orbifold action corresponding to the gauge group $\Gamma=\Z_N$,
\footnote{In the case of $X$ being a sum of line bundles with charges $\displaystyle q_i=({n_1\over d_1},{n_2\over d_2}\dots)$, $N$ will be given by $\lcm(d_1,d_2,\dots)$. }
and we quotient the theory by $\Z_2\ltimes\Z_N\cong \Z_{2N}$. Therefore we will need to account for
$2N$ sectors twisted by $\exp[\pi iJ]^k, k=0,\dots,2N-1$, where $J$ correspond to the $\GU(1)_L$ symmetry. 
CPT invariance exchanges the $k$-th and the $(2k-N)$-th sectors so we can restrict our analysis to the $0,1,\dots,N$ sectors. 
The massless spectrum will contain the universal sector, consisting of the fermonic degrees of freedom of the $N=1$ supergravity and axio-dilaton multiplets, together with the gauginos
of the hidden $\GE_8$, as well as some model-dependent vector and chiral multiplets. 
The $\GE_6$ neutral states, corresponding to first order deformations of the internal theory, are of main interest in our analysis and can only arise 
in the (NS,R) sector.

Let us recall, as pointed out in the case of LGOs \cite{Kachru:1993pg}, that
\begin{align}
\{\bQb,\bQb^{\dagger} \} = 2\Lb_0,
\end{align}
and hence the kernel of $\Lb_0$ is isomorphic to the cohomology of $\bQb$, which turns out to be computationally easier to handle.
Unfortunately, we do not have a similar technique at our disposal to describe the zero left-energy spectrum, and this condition has to be imposed by hand.
This, however, turns out to be quite possible since we can realize a $N=2$ left-moving algebra commuting with $\bQb$.
We can then work out the $\bQb$-cohomology at fixed $J_0=\bq$ and $L_0=E=0$.

\subsection*{$N=2$ left-moving algebra in $\bQb$-cohomology}

We showed above that the right-moving ground states are identified with $\bQb$-cohomology. Following the argument of 
\cite{Silverstein:1994ih} we can find representatives of an $N=2$ left-moving superconformal conformal algebra commuting with $\bQb$.
The currents are given by
\begin{align}
\label{eq:leftSCA}
\cJ_L &= \cX^\beta(D_\beta V^\alpha -\delta^\alpha_\beta)\cXb_{\alpha}- V^\alpha g_{\alpha\betab} \pz\Yb^{\betab}~,\nonumber\\
\cT  &=  -\pz Y^\alpha \left[g_{\alpha\betab}\pz\Yb^{\betab} -  g_{\gamma\betab,\alpha} \cX^\gamma \cXb^{\betab}\right] - \cX^\alpha\pz\cXb_\alpha- \half\pz\cJ_L~,\nonumber\\
\cG^+  &= i\sqrt{2} \left[ \cXb_{\alpha} \pz Y^\alpha - \pz(\cXb_{\alpha} V^\alpha)\right],\qquad
\cG^- = i\sqrt{2} \left[ \cX^\alpha g_{\alpha\betab} \pz\Yb^{\betab}\right]~
\end{align}
where $\cJ_L$ is identified with the $\GU(1)_L$ symmetry, $\cT$ is the stress-energy tensor and the remaining 
currents correspond to the supercharges.
These multiplets are $\cDb$-closed \footnote{Recall that since $\cDb$ and $\bQb$ are conjugate operators, we can study $\cDb$-cohomology instead of $\bQb$-cohomology.}
and their lowest components give operators whose action is well-defined in $\bQb$-cohomology. 
The OPEs of the left-moving fields are determined by the action \eqref{eq:compaction} and are given by
\begin{align}
\label{eq:OPEs}
y^\alpha(z) \rho_{\beta}(w) \sim \frac{1}{z-w}\delta^\alpha_\beta~,\qquad
\chi^\alpha(z)\chib_{\beta}(w) \sim \frac{1}{z-w}\delta^\alpha_\beta~.
\end{align}
It is easy to compute the central charge of the left-moving algebra 
\begin{align}
c = 3d+3\sum_{i=1}^n(1-2q_i)~,
\end{align}
where we recognize immediately the two contributions from the base and the LG fiber theories.
In table 1 
we have listed weights and charges of the fields, where $\bqb$ is the charge of $\GU(1)_R$.
\begin{table}[t]
\begin{center}
\label{table:charges}
\begin{tabular}{|c|c|c|c|c|c|c|c|c|}
\hline
~ 	
&$y^I$			&$\rho_I$				&$\chi^I$			&$\chib_I$	
&$\phi^i$			&$\rho_i$				&$\chi^i$			&$\chib_i$  \\ \hline
$\bq$	
&$0$		&$0$		&$-1$	&$1$
&$q_i$		&$-q_i$			&$q_i-1$			&$1-q_i$	\\ \hline
$2h$
&$0$		&$2$		&$1$		&$1$	
&$q_i$	&$2-q_i$	&$1+q_i$ 	& $1-q_i$ \\ \hline
$\bqb$ 
&$0$			&$0$				&$0$				&$0$
&$q_i$		&$-q_i$			&$q_i$			&$-q_i$ 
\\ \hline
\end{tabular}
\caption{Weights and charges of the fields.}
\end{center}
\end{table} 
We have then reduced the problem to a curved $bc-\beta\gamma$ system. That is, let $\{\mathfrak{U}_a\}$ be an open cover of $\bY$, then in each patch 
we have obtained a realization of the model as a free field theory. 
However, the transformation properties of the fields between patches fully represent the non-trivial geometry. 
The fields will thus transform as sections of appropriate bundles over $\bY$.

\subsection*{Twisted sectors}

In this section we will provide expressions for $E$, $\bq$ and $\bqb$ of the fields and the vacuum $|k\ra$ in the various twisted sectors.
For ease of exposition we will restrict to the case of $X$ being a sum of line bundles, but the formulae below can be extended to a more general $X$.
We choose the vacuum to be annihilated by all positive modes.
The moding of the left-moving fields in the $k$-th twisted sector are given by 
\begin{align}
y^\alpha(z)&=\sum_{r\in\Z-\nu_\alpha} y^\alpha_r z^{-r-h_\alpha} ~,   &\chi^\alpha(z)&=\sum_{r\in\Z-\nut_\alpha} \chi^\alpha_r z^{-r-\htt_\alpha}, \nonumber\\
\rho_\alpha(z)&=\sum_{r\in\Z+\nu_\alpha} \rho_{\alpha r} z^{-r+h_\alpha-1} ~,    &\chib_\alpha(z)&=\sum_{r\in\Z-\nut_\alpha} \chib_{\alpha r} z^{-r+\htt_\alpha-1},
\end{align}
where
\begin{align}
\nu_\alpha&={kq_\alpha\over2} \mod1~, &\nut_\alpha&={k(q_\alpha-1)\over2} \mod1~,  &h_\alpha&={q_\alpha\over2}~,  &\htt_\alpha&={q_\alpha+1\over2}~.
\end{align}
Also, we define $0\leq \nu_\alpha<1$ and $-1<\nut_\alpha\leq 0$.  
The (anti)commutation relations for the modes follow from the OPEs \eqref{eq:OPEs} and the quantum numbers of
the twisted vacuum $|k\ra$ are obtained by computing the 1-point functions of $T$ and $J_L$:
\begin{align}
\bq^{|k\ra}&=\sum_\alpha \left[ (q_\alpha-1)(\nut_\alpha+\half)-q_\alpha(\nut_\alpha-\half)   \right] ~, \nonumber\\
\bqb^{|k\ra}&=\sum_\alpha \left[ q_\alpha(\nut_\alpha+\half)+(q_\alpha-1)(-\nut_\alpha+\half)   \right] ~ , \nonumber\\
E^{|k\ra} &=\begin{cases}
0 \quad & \text{ for $k$ even}\\
 -{5\over8} + \half \sum_\alpha \left[ \nu_\alpha(1-\nu_\alpha)+\nut_\alpha(1+\nut_\alpha)   \right]   &  \text{ for $k$ odd}
\end{cases}.
\end{align}

In general, in order for $|k\ra$ to be a well-defined state, it must be accompanied by a wave-function over the space of bosonic zero modes; more precisely, 
$|k\ra$ transforms as a section of the holomorphic bundle over $B$ given by 
\begin{align}\label{bundlegroundst}
L_{|k\ra}=\begin{cases} \otimes_i L_i^{(\nut_i-\nu_i)}  \qquad\qquad\;  \text{ for $k$ even} \\
 \otimes_i L_i^{(\nut_i-\nu_i+\half)}  \qquad\quad\, \text{ for $k$ odd}
\end{cases}~.
\end{align}
Here we used the fact that the canonical bundle of $\bY$ is trivial. In fact, note that $\nu_\alpha-\nut_\alpha\in \Z$ for $k$ even and $\nu_\alpha-\nut_\alpha\in \Z+\half$ for $k$ odd; 
thus $L_{|k\ra}$ is indeed well-defined.

\subsection*{$\bQb$-cohomology and the spectral sequence}

Using the fact that the left-moving $N=2$ SCA in \eqref{eq:leftSCA} commutes with $\bQb$, we can compute its cohomology at fixed $\bq, E$, denoting the corresponding Hilbert space $\cH\equiv \cH|_{\bq,E}$.

As in the case of LGO, $\cH$ has a natural grading given by $\bqb$, $\cH=\oplus_{\bqb\in\Z+\half}\cH_{\bqb}$, and $\bQb$ acts as a differential, $\bQb : \cH_{\bqb} \rightarrow \cH_{\bqb+1}$.
Now, we would like to introduce an additional grading defined by the operator $U$ which assigns charge 1 to $\etab_\alpha$ and $-1$ to to $\eta^\alpha$.
Clearly, we have $[\bQb_0,U]=\bQb_0$ and $[\bQb_W,U]=0$. 
We then obtain a double grading on $\cH$ given by $U$ and $p=\bqb-U$, so that the operators
\begin{align}
\bQb_0: \cH_P^{p,U} \rightarrow \cH_P^{p,U+1}, \qquad \bQb_W : \cH_P^{p,U} \rightarrow \cH_P^{p+1,U}
\end{align}
act as the vertical and horizontal differentials. 
Recalling that $\{ \bQb_0,\bQb_W\}=0$, it follows that the cohomology of $\bQb=\bQb_0+\bQb_W$ is thus computed by a spectral sequence with first two stages 
\begin{align}
E_1^{p,U}=H^U_{\bQb_0}(\cH^{p,\bullet})~,\qquad 
E_2^{p,U}=H^p_{\bQb_W}H^U_{\bQb_0}(\cH^{\bullet,\bullet})~.
\end{align}
 The spectral sequence will always converge since $0\leq U \leq d = \dim B$ implies $d_{r>d} =0$  and 
$E_{d +1}^{p,U}=E_\infty^{p,U}=H^{p,U}_{\bQb}(\cH^{\bullet,\bullet})$.
A problem might arise by the non-compactness of the $W=0$ geometry: the first step of the spectral sequence is
$\bQb_0$-cohomology which is equivalent to horizontal Doulbeaut cohomology, and this will in general give
infinite dimensional vector spaces even at fixed $E$ and $\bq$.
 In order to have a well-defined counting we remember that the
theory at $W=0$ has an extra $\GU(1)^{\oplus n}$ symmetry which rotates each of the $\phi$ separately. Therefore
the fine grading $\br=(r_1,\dots,r_n)\in \Z^n$ associated to the monomial $\prod_{i=1}^n \phi_i^{r_i}$ is a refinement of the 
coarse grading given by $E,\bq$, and it yields a well-defined counting problem which can be phrased in terms of sheaf cohomology over $B$.

The second step of the spectral sequence is $\bQb_W$-cohomology and higher differentials will be determined in terms of  $\bQb_0$ and $\bQb_W$
by the standard zig-zag argument \cite{Bott:1982df}.

We will not dwell here on the geometrical description of each twisted sector, but we will just summarize the main features.
At $k=1$ the geometry is given by the full $\bY$, and the states are organized as horizontal forms valued in the holomorphic bundle
\begin{align}
\label{eq:Bdef}
B_{s,t,q} \equiv \wedge^s T_{\bY} \otimes \wedge^t T^\ast_{\bY} \otimes \Sym^q (T_{\bY})~.
\end{align}
For odd $k>1$ the geometry is given by a sector-dependent sub-bundle ${\bY}_k$ of $\bY$, determined essentially by
the fact that the vacuum $|k\ra$ transforms as a non-trivial holomorphic section over $B$.
For $k$ even, the left moving energy $E_{|k\ra}=0$ and we just restrict to zero-modes for all fields.

\section{An example: the octic in $\P^4_{\{2,2,2,1,1\}}$}

Let us start by briefly summarizing the findings of \cite{Aspinwall:2010ve} for the resolved octic hypersurface in the weighted projective space $\P^4_{\{2,2,2,1,1\}}$. 
In the large radius phase, the counting for $h^1(\End T)$ for generic complex structure is 188 while it get enhanced to 200 at the Fermat locus. 
However, the GLSM deformations, given by the bottom row of the spectral sequence in the language of \cite{Aspinwall:2010ve},
are just 179, and there is strong evidence that they correspond to exactly marginal deformations of the theory \cite{Kreuzer:2010ph,Basu:2003bq}. 
That means that there is room for instanton-induced masses for the extra 9 generic singlets. The results of the spectrum analysis at both the LG point and in the orbifold phase agree with
the large radius counting and thus seem to exclude this possibility. 

Here, in some sense we are going to complete their analysis by providing the counting for massless gauge singlets in the hybrid phase,
giving a sketch of the application of methods developed in \cite{Bertolini:2013xga}.
The hybrid model is a phase of the two parameter GLSM \cite{Candelas:1993dm, Morrison:1994fr},
with geometry given by $\cO(-2)\oplus\cO^{\oplus3}\rightarrow\P^1$ and superpotential 
\begin{align}
W = \sum_{i=1}^{4} F_{[4-i]}(\phi^1)^i~,
\end{align}
where $F_{[d]}$ is a generic degree $d$ polynomial in $\phi^j$, $j=2,3,4$, with coefficients in $H^0(\P_1,\cO(8-2d))$.
The quantum numbers of the ground states of the twisted sectors, as well as charges of the fiber fields are given in table 2. 
In this example $\Pic B = H^2(\P^1,\Z)=\Z$, and the dual bundle $L_{|k\ra}^\ast$ is simply determined by the line bundle $\cO(\ell_k)$ over $\P^1$.
\begin{table}[!t]
\renewcommand{\arraystretch}{1.3}
\begin{center}
\begin{tabular}{cc}
\begin{tabular}{|c|c|c|c|c|c|c|c|c|c|c|c|}
\hline
$k$ 		&$E^{|k\ra}$		&$\bq^{|k\ra}$		&$\bqb^{|k\ra}$   	&$\ell_k$	 	&$\nu_{i}$		&$\nut_{i}$	         &$\nu_I$			&$\nut_I$			\\ \hline
$0$		&$0$			&$-\ff{3}{2}$		&$-\ff{3}{2}$		&$0$		&$0$			&$0$  			&$0$	  		&$0$			\\ \hline
$1$		&$-1$			&$0$			&$-\ff{3}{2}$       	&$0$      		&$\ff{1}{8}$		&$-\ff{3}{8}$	 	&$0$			&$-\ff{1}{2}$		\\ \hline
$2$		&$0$			&$\ff{1}{2}$		&$-\ff{3}{2}$  		&$-2$		&$\ff{1}{4}$		&$-\ff{3}{4}$ 		&$0$			&$0$			\\ \hline
$3$		&$-\ff{1}{2}$		&$-1$			&$-\half$			&$0$		&$\ff{3}{8}$		&$-\ff{1}{8}$	 	&$0$			&$-\ff{1}{2}$		\\ \hline
$4$		&$0$			&$-\half$			&$-\half$			&$-2$		&$\ff{1}{2}$		&$-\ff{1}{2}$	 	&$0$			&$0$		    	\\ \hline
\end{tabular}
&
\begin{tabular}{|c|c|c|c|c|}
\hline
~ 		&$\phi^i$			&$\rho_i$				&$\chi^i$			&$\chib_i$  \\ \hline
$\bq$	&$\ff{1}{4}$		&$-\ff{1}{4}$			&$-\ff{3}{4}$			&$\ff{3}{4}$	\\ \hline
$\bqb$ &$\ff{1}{4}$		&$-\ff{1}{4}$			&$\ff{1}{4}$			&$-\ff{1}{4}$ \\ \hline
\end{tabular}
\end{tabular}
\end{center}
\label{table:octic}
\caption{Quantum numbers for the octic model.}
\end{table}
We will concentrate on $\GE_6$-singlet states, and therefore we are going to restrict our analysis to odd twisted sectors, i.e. $k=1,3$ and $E=\bq=0$.

In the $k=1$ sector, the first stage of the spectral sequence is obtained by taking $\bQb_0$-cohomology on the operators at $E=\bq=0$. We obtain 
\begin{equation}\label{eq:spectrseqE1octic}
\begin{matrix}\vspace{15mm}\\E_1^{p,u}:\end{matrix}
\begin{xy}
\xymatrix@C=12mm@R=5mm{
{\begin{matrix}
H^1\left(\bY,B_{0,0,1} \right)_{3}
\\\oplus\\ 
H^1\left(\bY, B_{1,1,0} \right)_{22}
\end{matrix}}   \ar[r]^{\bQb_W}  &
H^1\left(\bY, B_{0,1,0} \right)_{15}   \\
{\begin{matrix}
H^0\left(\bY,B_{0,0,1}\right)_{23}
\\\oplus\\ 
H^0\left(\bY,B_{1,1,0} \right)_{21}
\end{matrix}}   \ar[r]^{\bQb_W}  
  &
{\begin{matrix}
H^0\left(\bY, B_{0,1,0}\right)_{290}
\end{matrix}}
}
\save="x"!LD+<-6mm,0pt>;"x"!RD+<40pt,0pt>**\dir{-}?>*\dir{>}\restore
\save="x"!LD+<70mm,-3mm>;"x"!LU+<70mm,2mm>**\dir{-}?>*\dir{>}\restore
\save!RD+<-55mm,-4mm>*{-\ff{3}{2}}\restore
\save!RD+<-18mm,-4mm>*{-\half}\restore
\save!RD+<13mm,-3mm>*{p}\restore
\save!CL+<67mm,18mm>*{U}\restore
\end{xy}
\end{equation}
The dimension of each group is indicated as a subscript.  The bottom row of \eqref{eq:spectrseqE1octic} is, in some sense, universal:
for generic $W$ the kernel is 1-dimensional, corresponding to the current associated to the $\GU(1)_L$ symmetry; for a more specific 
form of $W$ we can increase the kernel of $\bQb_W$ and obtain enhanced symmetry. The cohomology of the first row is model-dependent 
and it can be shown that $\bQb_W$ is surjective for generic $W$ while $\coker \bQb_W=6$ at the Fermat form. 

In the $k=3$ sector the situation is again straightforward because the vacuum $|3\ra$ transforms trivially, and the
geometry is still provided by the full $\bY$
\begin{equation}\label{eq:spectrseqE3octic}
\begin{matrix}\vspace{10mm}\\E_1^{p,u}:\end{matrix}
\begin{xy}
\xymatrix@C=12mm@R=5mm{
0 & H^1\left(\bY, B_{0,1,0} \right)_{15}   \\
 0 & {\begin{matrix}
H^0\left(\bY, B_{0,1,0}\right)_{290}
\end{matrix}}
}
\save="x"!LD+<-6mm,0pt>;"x"!RD+<40pt,0pt>**\dir{-}?>*\dir{>}\restore
\save="x"!LD+<47mm,-3mm>;"x"!LU+<47mm,2mm>**\dir{-}?>*\dir{>}\restore
\save!RD+<-43mm,-4mm>*{-\ff{3}{2}}\restore
\save!RD+<-15mm,-4mm>*{-\half}\restore
\save!RD+<13mm,-3mm>*{p}\restore
\save!CL+<44mm,12mm>*{U}\restore
\end{xy}
\end{equation}
Moreover, $\bQb_W$ is identically zero on states at $E=\bq=0$, 
and the spectral sequence degenerates already at the first stage. We thus count 25 massless $\GE_6$-singlets. 

We can thus summarize our results: we find 282 singlets at generic $W$, while the number increases to 297 for $W$ at the Fermat form. 
This value differs by one from the result at the LG point. There, the ``extra" 6 singlets with respect to the value of 188 at large radius where shown, by using
mirror symmetry, to acquire a K\"ahler-dependent mass. 
It is therefore natural to try the same approach in the hybrid limit.

\subsection*{A little mirror symmetry}

The Gepner model is a sum of minimal models $A_3^{\oplus3}\oplus A_7^{\oplus2}$, and the mirror model is obtained by taking a $\Z_3^{\oplus3}$ orbifold. 
The superpotential for the mirror model is 
\begin{align}
\hat W = \phi_1^4 + \phi_2^4 + \phi_3^4 + \phi_4^8 + \phi_5^8 -8\psi \phi_1\phi_2\phi_3\phi_4\phi_5 -4 \chi \phi_4^4\phi_5^4~.
\end{align}
The Gepner model corresponds to $\psi=\chi=0$, while the hybrid limit is reached at $\psi=0,\chi\rightarrow\infty$.
Since we are interested in states in the untwisted sector $k=1$, we do not need to go in the details of the orbifold. 
In particular, we want to see what happens if we turn is a $\chi$ deformation. At $\bq=0$ we have
\begin{equation}\label{eq:k1Gepnermirror}
\begin{xy}
\xymatrix@C=20mm@R=2mm{
  { \begin{matrix}
 G^{ia}_{[1]} \gamma_i\gammab_a |1\ra_{12} \oplus   \gamma_i\gammab_j |1\ra_9 \oplus   \gamma_a\gammab_b |1\ra_4 \\\oplus\\
G^i_{[2]}  \rho_i|1\ra_{18} \oplus G^a_{[1]}  \rho_a|1\ra_4 
 \end{matrix} }
   \ar[r]^-{\bQb_W} &
   {\begin{matrix}
G^i_{[6]} \gamma_i |1\ra_{150} \\\oplus\\
G^a_{[7]} \gamma_a |1\ra_{140}
\end{matrix}}\\
   \bqb=-\half &    \bqb=\half
}
\end{xy}
\end{equation}
where $i,j=1,2,3$, $a,b=4,5$, and $G_{[d]}$ is a generic degree $d$ polynomial in $\phi_i$ and $\phi_a$ with weights 2 and 1, respectively.
In this sector we have $\bQb_W = \gammab^\dag_\alpha W_\alpha + \gamma^\alpha W_{\alpha\beta}\rho^\dag_\beta$, $\alpha,\beta = 1,\dots, 5$, which reads
\begin{align}
\bQb_W &= 4 \gammab^\dag_i \phi_i^3 +8 \gammab^\dag_a ( \phi_a^7 -2 \chi \phi_a^3\phi_b^4)   + 12\gamma^i \phi_i^2\rho^\dag_i + 
12 \gamma^i\phi_i^2\rho_i^\dag \nonumber\\
&\quad + 8\gamma^a ( 7 \phi_a^6 - 6 \chi \phi_a^2\phi_b^4 ) \rho_a^\dag + 64\chi \gamma^a \phi_a^3\phi_b^3 \rho_b^\dag~.
\end{align}
At the Gepner point $\ker \bQb_W$ is 5-dimensional given by $(\phi_\alpha\rho_\alpha+ \gamma_\alpha \gammab_\alpha)|1\ra$. 
Clearly, a $\chi$ deformation leaves unaffected the states for $\alpha=i$ but removes the degeneracy for $\alpha=a$. 
The the sum of such states is never lifted and we have that exactly one vector-singlet pair is lifted by a $\chi$ deformation.
This explains the discrepancy in the number of singlets found in the hybrid limit ($\chi\rightarrow\infty$) with respect to the value at the Gepner point.

\bibliographystyle{amsplain}

\bibliography{./bigref}

\end{document}